\documentclass[12pt,oneside]{article}

\usepackage{amssymb}
\usepackage{amsxtra}
\usepackage{amsmath}
\usepackage{amsfonts}
\usepackage{CJK}
\usepackage[titletoc]{appendix}
\usepackage{graphicx}

\numberwithin{equation}{section}

\newcommand{\eqa}{\begin{eqnarray}}
\newcommand{\eeqa}{\end{eqnarray}}
\newcommand{\beq}{\begin{equation}}
\newcommand{\eeq}{\end{equation}}
\newcommand{\nn}{\nonumber}
\newcommand{\p}{\partial}

\allowdisplaybreaks

\begin{document}

\date{}
\author{Beibei Hu$^{a,b,}$\thanks{Corresponding author. E-mail address: hu\_chzu@163.com(B.-b. Hu).} , Wen-Xiu Ma$^{c,d,e}$, Tiecheng Xia$^{b}$, Ling Zhang$^{a}$ \\
\small \textit{a.School of Mathematicas and Finance, Chuzhou University, Anhui, 239000, China}\\
\small \textit{b.Department of Mathematics, Shanghai University, Shanghai 200444, China}\\
\small \textit{c.Department of Mathematics and Statistics, University of South Florida, Tampa, FL, 33620-5700, USA}\\
\small \textit{d.College of Mathematics and Systems Science, Shandong University }\\
\small \textit{of Science and Technology, Qingdao 266590, Shandong, China}\\
\small \textit{e.International Institute for Symmetry Analysis and Mathematical Modeling, Department of Mathematical}\\
\small \textit{ Sciences, North-West University, Mafikeng Campus, Private Bag X2046, Mmabatho 2735, South Africa}
}
\title{Nonlinear integrable couplings of a generalized super Ablowitz-Kaup-Newell-Segur hierarchy and its super bi-Hamiltonian structures
}
\maketitle

\begin{abstract}

In this paper, a new generalized $5\times5$ matrix spectral problem of Ablowitz-Kaup-Newell-Segur(AKNS) type associated with
the enlarged matrix Lie super algebra is proposed and its corresponding super soliton hierarchy is established.
The super variational identities is used to furnish super-Hamiltonian structures for the resulting super soliton hierarchy.
\end{abstract}

\section{Introduction}

\quad\;\;It is known that super integrable systems provide interesting and important  models in the supersymmetry theory
Supersymmetry is originated in 1970s when physicists have proposed simple models with
supersymmetric colors in string models and mathematical physics respectively. After that,
Wess and Zumino [1] applied supersymmetry to the four-dimensional spacetime. Unfortunately,
the supersymmetry partners of any particle have not been found so far, and it is generally
believed that this symmetry is spontaneous rupture. In order to unify two kinds of particles
with different spin and statistical properties-Boson and Fermion, theoretical
physicists proposed the concept of hyperspace in the study of unified field theory and quantum
field theory. Inspired by this, mathematicians developed the super analysis, the hypergeometric and the super algebra.

Due to the importance of supersymmetry in physics(especially in the exploration of the relationship
between supersymmetric conformal field and chord theory), which has attracted great attention
for the study of super integrable systems associated with Lie super algebra, many classical
solition equations have been extended to be the super completely integrable system. For examples,
the super Ablowitz-Kaup-Newell-Segur(AKNS) hierarchy [2-10], the super Dirac hierarchy [4,11-14], the super Kaup-Newell(KN)
hierarchy [14-18], and others [19-28].
Among those, Hu \cite{hu1997} and Ma \cite{ma2008} has made a
great contribution. Hu \cite {hu1997} proposed the super-trace identity, which is an effective
tool to constructing super Hamiltonian structures of super integrable equations.
In 2008, Ma given the proof of the super-trace identity and the super Hamiltonian structure
of many super integrable equations is established by the super-trace identity \cite{ma2008,ma2010}.
Meanwhile, constructing nonlinear superintegrable couplings
with enlarging matrix Lie super algebra is one of pretty interesting topics in supermodel
theory [32-38]. There are much richer mathematical structures behind nonlinear super
integrable couplings than scalar superintegrable equations. Moreover, the study of
superintegrable couplings generalizes the classical integrable couplings theory and
provides clues toward complete classification of superintegrable equations [39-45].

In Ref.\cite{You2011}, You considered an enlarged super AKNS matrix spectral
problem is given by
\eqa {\small \phi_x=M\phi,
  M=\left(\begin{array}{ccccc}
\lambda & p & 0 & r& \alpha\\
q&-\lambda&s&0&\beta\\
0&0&\lambda&p+r&0\\
0&0&q+s&-\lambda&0\\
\beta &-\alpha &-\beta &\alpha &0
\end{array} \right),
\phi=\left(\begin{array}{c}
\phi_1\\
\phi_2\\
\phi_3\\
\phi_4\\
\phi_5\end{array} \right),
u=\left(\begin{array}{c}
p\\
q\\
\alpha\\
\beta\\
r\\
s\end{array} \right)},\label{1.1}\eeqa
where $\lambda$ is the spectral parameter, $p,q,r$ and $s$ are even potentials, but $\alpha$ and $\beta$ are odd ones.
Take $\alpha=\beta=0$, the hierarchy \eqref{1.1} reduces to a nonlinear integrable couplings of the classical AKNS hierarchy \cite{MA2010}.
Whose super Hamiltonian structure is furnished by super trace identity.
Recently, Shen et al \cite{Shen2017}. considered a generalized spatial spectral problem of AKNS integrable
coupling as follows
\eqa {\small \phi_x=U\phi, U=\left(\begin{array}{cccc}
\lambda+\omega & p & 0 & r\\
q&-\lambda-\omega&s&0\\
0&0&\lambda+\omega&p\\
0&0&q&-\lambda-\omega
\end{array} \right), \phi=\left(\begin{array}{c}
\phi_1\\
\phi_2\\
\phi_3\\
\phi_4\end{array} \right), u=\left(\begin{array}{c}
p\\
q\\
r\\
s\end{array} \right),}\label{1.2}\eeqa
where $\omega=\epsilon(ps+qr)$ and $\lambda$ is the spectral parameter, $p,q,r$ and $s$ are commuting variables.
Obviously, when $\epsilon=0$, this generalized spatial spectral problem \eqref{1.2} is reduced to a new case
of AKNS integrable couplings \cite{Shen2014}. Whose bi-Hamiltonian structures were constructed by using the component-trace
identity in \cite{Shen2017}. Inspired by those spatial spectral problem, in this paper, we would like to
construct nonlinear super integrable couplings of a generalized super AKNS hirearchy.

The rest of this paper is organized as follows. In Section 2, we will enlarge the Lie superalgebra $sl(2,1)$
to the Lie superalgebra $sl(4,1)$. In Section 3, we will construct a generalization
of the super AKNS integrable coupling hierarchy from zero curvature equations, based on the above-mentioned
generalized spatial spectral problem \eqref{1.1}. In Section 4, the super bi-Hamiltonian form will
be presented for the obtained super integrable couplings of the generalized super AKNS hierarchy by making use of the super
trace identity. For the sake of convenience, we will use the mathematical software Maple to deal with some
complicated symbolic computations. And the last section is devoted to conclusions and discussions.

\section{Enlargement of a Lie Superalgebra}

\quad\;\;In this section, we consider the Lie superalgebra $sl(2,1)$. Its basis is
\eqa &&
E_1=\left(\begin{array}{ccc}
1&0&0\\
0&-1&0\\
0&0&0\end{array} \right),
E_2=\left(\begin{array}{ccc}
0&1&0\\
0&0&0\\
0&0&0\end{array} \right),
E_3=\left(\begin{array}{ccc}
0&0&0\\
1&0&0\\
0&0&0\end{array} \right),\nn\\
&&E_4=\left(\begin{array}{ccc}
0&0&1\\
0&0&0\\
0&-1&0\end{array} \right),
E_5=\left(\begin{array}{ccc}
0&0&0\\
0&0&1\\
1&0&0\end{array} \right).\nn\eeqa
where $E_1,E_2,E_3$ are even elements and $E_4,E_5$ are odd ones,
$[.,.]$ and $[.,.]_{+}$ denote the commutator and the anticommutator,
satisfy the following operational relations:
\eqa&&
[E_1,E_2]=2E_2,[E_1,E_3]=-2E_3,[E_2,E_3]=E_1,\nn\\
&&[E_1,E_4]=[E_2,E_5]=E_4,[E_1,E_5]=[E_4,E_3]=-E_5,[E_2,E_4]=[E_3,E_5]=0,\nn\\
&&[E_4,E_4]_{+}=-2E_2,[E_5,E_5]_{+}=2E_3,[E_4,E_5]_{+}=[E_5,E_4]_{+}=E_1. \label{2.1}\eeqa

Let us enlarge the Lie superalgebra $sl(2,1)$ to the Lie
superalgebra $sl(4,1)$ with a basis
\eqa &&
e_1=\left(\begin{array}{ccccc}
1&0&0&0&0\\
0&-1&0&0&0\\
0&0&1&0&0\\
0&0&0&-1&0\\
0&0&0&0&0\end{array} \right),
e_2=\left(\begin{array}{ccccc}
0&1&0&0&0\\
0&0&0&0&0\\
0&0&0&1&0\\
0&0&0&0&0\\
0&0&0&0&0\end{array} \right),
e_3=\left(\begin{array}{ccccc}
0&0&0&0&0\\
1&0&0&0&0\\
0&0&0&0&0\\
0&0&1&0&0\\
0&0&0&0&0\end{array} \right),\nn\\&&
e_4=\left(\begin{array}{ccccc}
0&0&1&0&0\\
0&0&0&-1&0\\
0&0&1&0&0\\
0&0&0&-1&0\\
0&0&0&0&0\end{array} \right),
e_5=\left(\begin{array}{ccccc}
0&0&0&1&0\\
0&0&0&0&0\\
0&0&0&1&0\\
0&0&0&0&0\\
0&0&0&0&0\end{array} \right)
e_6=\left(\begin{array}{ccccc}
0&0&0&0&0\\
0&0&1&0&0\\
0&0&0&0&0\\
0&0&1&0&0\\
0&0&0&0&0\end{array} \right)\nn\\&&
e_7=\left(\begin{array}{ccccc}
0&0&0&0&1\\
0&0&0&0&0\\
0&0&0&0&0\\
0&0&0&0&0\\
0&-1&0&1&0\end{array} \right)
e_8=\left(\begin{array}{ccccc}
0&0&0&0&0\\
0&0&0&0&1\\
0&0&0&0&0\\
0&0&0&0&0\\
1&0&-1&0&0\end{array} \right)
.\nn\eeqa
where $e_1,e_2,e_3,e_4,e_5,e_6$ are even and $e_7,e_8$ are odd,
$[.,.]$ and $[.,.]_{+}$ denote the commutator and the anticommutator.
The generators of the Lie superalgebra $sl(4,1),e_i,0\leq i\leq8$, satisfy
the following operational relations:
\eqa &&[e_1,e_2]=2e_2,[e_1,e_3]=-2e_3,[e_1,e_5]=-[e_2,e_4]=[e_4,e_5]=2e_5,[e_2,e_3]=e_1,\nn\\&&
[e_1,e_6]=-[e_3,e_4]=[e_4,e_6]=-2e_6,[e_1,e_7]=[e_2,e_8]=e_7,[e_1,e_8]=-[e_3,e_7]=-e_8,\nn\\&&
[e_2,e_6]=-[e_3,e_5]=[e_5,e_6]=e_4,
[e_1,e_4]=[e_2,e_5]=[e_2,e_7]=[e_3,e_6]=[e_3,e_8]=0,\nn\\&&
[e_4,e_7]=[e_4,e_8]=[e_5,e_7]=[e_5,e_8]=[e_6,e_7]=[e_6,e_8]=0\nn\\&&
[e_7,e_8]_{+}=e_1-e_4,[e_7,e_7]_{+}=2e_5-2e_2,[e_8,e_8]_{+}=2e_3-2e_6
. \label{2.2}\eeqa
Define a loop super algebra corresponding to the Lie super algebra
$sl(4,1)$, and denote by
\beq sl(4,1)=sl(4,1)\otimes[\lambda,\lambda^{-1}]
={e_i\lambda^m,e_i\in sl(4,1),i=1,2,\ldots,8;m=0,\pm1,\pm2,\ldots}.\label{2.3}\eeq
The corresponding (anti)commutative relations are given as
\beq [e_i\lambda^m,e_j\lambda^n]=[e_i,e_j]\lambda^{m+n},\forall e_i,e_j\in sl(4,1).\label{2.4}\eeq

\section{ Nonlinear generalized super integrable couplings of the super AKNS hierarchy}

\quad\;\;In this section, we shall construct nonlinear integrable couplings of a
generalized super AKNS hierarchy from an enlarging matrix Lie super algebra.
Consider the following spatial spectral problem
 \eqa {\small \phi_x=M\phi,
  M=\left(\begin{array}{ccccc}
\lambda+h & p & 0 & r& \alpha\\
q&-\lambda-h&s&0&\beta\\
0&0&\lambda+h&p+r&0\\
0&0&q+s&-\lambda-h&0\\
\beta &-\alpha &-\beta &\alpha &0
\end{array} \right),
\phi=\left(\begin{array}{c}
\phi_1\\
\phi_2\\
\phi_3\\
\phi_4\\
\phi_5\end{array} \right),
u=\left(\begin{array}{c}
p\\
q\\
\alpha\\
\beta\\
r\\
s\end{array} \right)},\label{3.1}\eeqa
where $h=\mu(ps+qr+rs-2\alpha\beta)$ with $\mu$ being an arbitrary even constant,
$\lambda$ is the spectral parameter, $p,q,r$ and $s$ are even potentials,
and $\alpha$ and $\beta$ are odd potentials. Obviously, the spatial spectral
problem \eqref{3.1} with $\mu=0$ reduces to the standard nonlinear integrable
couplings of super AKNS hierarchy case \cite{You2011}.

In order to derive super integrable couplings of a generalized super integrable hierarchy
associated with the spatial spectral problem \eqref{3.1}, we solve the stationary zero curvature equation
\beq N_x=[M,N].\label{3.2}\eeq
where
\eqa {\small N=\left(\begin{array}{ccccc}
A&B&E&F&\rho\\
C&-A&G&-E&\delta\\
0&0&A+E&B+F&0\\
0&0&C+G&-A-E&0\\
\delta&-\rho&-\delta&\rho&0\end{array} \right)},\label{3.3}\eeqa
in which the corresponding $A,B,C,E,F,G$ are even elements and $\rho,\delta$, are odd elements.

Substituting $M$ in \eqref{3.1} and $N$ in \eqref{3.3} into Eq.\eqref{3.2} yields
\eqa
\left\{\begin{array}{l}
A_{x}=pC-qB-\alpha\delta+\beta\rho,\\
B_{x}=2\lambda B-2pA-2\alpha\rho+2hB,\\
C_{x}=-2\lambda C+2qA+2\beta\delta-2hC,\\
E_{x}=(p+r)G+rC-sB-(q+s)F-\alpha\delta-\beta\rho,\\
F_{x}=2\lambda F-2(p+r)E-2rA+2\alpha\rho+2hF,\\
G_{x}=-2\lambda G+2(q+s)E+2sA-2\beta\delta-2hG,\\
\rho_{x}=\lambda\rho+p\delta-\alpha A-\beta B+h\rho,\\
\delta_{x}=-\lambda\delta+\beta A-\alpha C+q\rho-h\delta.\\
\end{array}\right.\label{3.4}\eeqa
Choosing
\eqa &&
A=\sum^n_{m\geq0}a_m\lambda^{-m},B=\sum^n_{m\geq0}b_m\lambda^{-m},
C=\sum^n_{m\geq0}c_m\lambda^{-m},E=\sum^n_{m\geq0}e_m\lambda^{-m},\nn\\&&
F=\sum^n_{m\geq0}f_m\lambda^{-m},G=\sum^n_{m\geq0}g_m\lambda^{-m},
\rho=\sum^n_{m\geq0}\rho_m\lambda^{-m},\delta=\sum^n_{m\geq0}\delta_m\lambda^{-m},\label{3.5}\eeqa
and comparing the coefficients of the same powers of $\lambda$ in Eq.\eqref{3.4}, we have
\eqa
\left\{\begin{array}{l}
a_{m,x}=pc_{m}-qb_{m}+\alpha\delta_{m}+\beta\rho_{m},\\
e_{m,x}=rc_{m}-sb_{m}+(p+r)g_m-(q+s)f_m-\alpha\delta_{m}-\beta\rho_{m},\\
b_{m+1}=\frac{1}{2}b_{m,x}+pa_{m}+\alpha\rho_{m}-hb_{m},\\
c_{m+1}=-\frac{1}{2}c_{m,x}+qa_{m}+\beta\delta_{m}-hc_{m},\\
f_{m+1}=\frac{1}{2}f_{m,x}+ra_{m}+(p+r)e_m-\alpha\rho_{m}-hf_{m},\\
g_{m+1}=-\frac{1}{2}g_{m,x}+sa_{m}+(q+s)e_m-\beta\delta_{m}-hg_{m},\\
\rho_{m+1}=\rho_{m,x}+\alpha a_{m}+\beta b_{m}-p\delta_{m}-h\rho_{m},\\
\delta_{m+1}=-\delta_{m,x}+\beta a_{m}-\alpha c_{m}+q\rho_{m}-h\delta_{m},\\
\end{array}\right.\label{3.6}\eeqa
which results in the recurrence relations
\eqa
\left\{\begin{array}{l}
(c_{m+1}, b_{m+1},\delta_{m+1}, \rho_{m+1}, g_{m+1}, f_{m+1})^T=L( c_{m},b_{m},\delta_{m},\rho_{m},g_{m},f_{m})^T,\\
a_m=\p^{-1}(rc_{m}-qb_{m}+\alpha\delta_{m}+\beta\rho_{m}),\\
e_m=\p^{-1}(rc_{m}-sb_{m}-\alpha\delta_{m}-\beta\rho_{m}+(p+r)g_m-(q+s)f_m),
\end{array}\right.\label{3.7}\eeqa
where the recursion operator $L$ has the following form
\beq
L=\left(\begin{array}{ccc}
L_1 & L_2 & 0 \\
L_3& L_4 & 0 \\
L_5 & -L_2 & L_6
\end{array} \right).\label{3.8}\eeq
with
\eqa &&
L_1=\left(\begin{array}{cc}
q\p^{-1}p-\frac{1}{2}\p-h & -q\p^{-1}q \\
p\p^{-1}p & -p\p^{-1}q+\frac{1}{2}\p-h
\end{array} \right),
L_2=\left(\begin{array}{cc}
q\p^{-1}\alpha+\beta & q\p^{-1}\beta \\
p\p^{-1}\alpha & p\p^{-1}\beta+\alpha
\end{array} \right),\nn\\&&
L_3=\left(\begin{array}{cc}
\beta\p^{-1}p-\alpha & -\beta\p^{-1}q \\
\alpha\p^{-1}p & -\alpha\p^{-1}q+\beta
\end{array} \right),
L_4=\left(\begin{array}{cc}
\beta\p^{-1}\alpha-\p-h & \beta\p^{-1}\beta+q \\
\alpha\p^{-1}\alpha-p & \alpha\p^{-1}\beta+\p-h
\end{array} \right),\nn\\&&
L_5=\left(\begin{array}{cc}
s\p^{-1}p+(q+s)\p^{-1}r & -s\p^{-1}q-(q+s)\p^{-1}s\\
r\p^{-1}p+(p+r)\p^{-1}r & -r\p^{-1}q-(p+r)\p^{-1}s
\end{array} \right),\nn\\&&
L_6=\left(\begin{array}{cc}
(q+s)\p^{-1}(p+r)-\frac{1}{2}\p-h & -(q+s)\p^{-1}(q+s) \\
(p+r)\p^{-1}(p+r) & -(p+r)\p^{-1}(q+s)+\frac{1}{2}\p-h
\end{array} \right).
\nn\eeqa

Upon choosing the initial conditions $a_0=e_0=1,b_0=c_0=e_0=f_0=g_0=\rho_0=\delta_0=0$,
all other $a_j,b_j,c_j,\rho_j,\delta_j$$(j\geq1)$ can be worked out uniquely by
the recurrence relations \eqref{3.5} and by using of symbolic computation
software(Maple). We list the first three sets as follows:
\eqa &&
a_1=e_1=0,b_1=p,c_1=q,f_1=p+r,g_1=q+s,\rho_1=\alpha,\delta_1=\beta,\nn\\&&
a_2=-\frac{1}{2}(pq+2\alpha\beta),
b_2=\frac{1}{2}p_x-hp,
c_2=-\frac{1}{2}q_x-hq,\nn\\&&
e_2=-(ps+qr+rs+\frac{1}{2}pq-\alpha\beta),
f_2=\frac{1}{2}p_x+\frac{1}{2}r_x-hp-hr,\nn\\&&
g_2=-\frac{1}{2}q_x-\frac{1}{2}s_x-hq-hs,
\rho_2=\alpha_x-h\alpha,\delta_2=-\beta_x-h\beta,
\nn\\&&
a_3=\frac{1}{4}(pq_x-p_xq)+\alpha\beta_x-\alpha_x\beta+h(pq+2\alpha\beta),\nn\\&&
b_3=\frac{1}{4}p_{xx}-\frac{1}{2}h_xp-hp_x-\frac{1}{2}(pq+2\alpha\beta)p+\alpha\alpha_x+h^2p,\nn\\&&
c_3=\frac{1}{4}q_{xx}+\frac{1}{2}h_xq+hq_x-\frac{1}{2}(pq+2\alpha\beta)q-\beta\beta_x+h^2q,\nn\\&&
e_3=\frac{1}{2}(ps_x-p_xs+q_x r-q r_x+rs_x-r_xs+\frac{1}{2}pq_x-\frac{1}{2}p_xq)\nn\\&&
\quad\quad\;-(\alpha\beta_x-\alpha_x\beta)+2h(ps+qr+rs+\frac{1}{2}pq-\alpha\beta),\nn\\&&
f_3=\frac{1}{4}p_{xx}+\frac{1}{2}r_{xx}-\frac{1}{2}h_x(p+2r)-hp_x-2hr_x-\frac{1}{2}(pq+2\alpha\beta)r\nn\\&&
\quad\quad\;-\alpha\alpha_x+h^2(p+2r)-(p+r)(\frac{1}{2}pq+ps+qr+rs-\alpha\beta),\nn\\&&
g_3=\frac{1}{4}q_{xx}+\frac{1}{2}s_{xx}+\frac{1}{2}h_x(q+2s)+hq_x+2hs_x-\frac{1}{2}(pq+2\alpha\beta)s\nn\\&&
\quad\quad\;+\beta\beta_x+h^2(q+2s)-(q+s)(\frac{1}{2}pq+ps+qr+rs-\alpha\beta),\nn\\&&
\rho_3=\alpha_{xx}-h_x\alpha-2h\alpha_x-\frac{1}{2}(pq+2\alpha\beta)\alpha+h^2\alpha+\frac{1}{2}p_x\beta+p\beta_x,\nn\\&&
\delta_3=\beta_{xx}+h_x\beta+2h\beta_x-\frac{1}{2}(pq+2\alpha\beta)\beta+h^2\beta+\frac{1}{2}q_x\alpha+q\alpha_x. \nn\eeqa
Let us consider the spectral problem \eqref{3.1} with the following auxiliary spectral problem:
$$\phi_{t_n}=N^{(n)}\phi$$
where
\beq N^{(n)}=N_{+}^{(n)}+\Delta_n=\sum^n_{m=0}\left(\begin{array}{ccccc}
a_m&b_m&e_m&f_m&\rho_m\\
c_m&-a_m&g_m&-e_m&\delta_m\\
0&0&a_m+e_m&b_m+f_m&0\\
0&0&c_m+g_m&-a_m-e_m&0\\
\delta_m&-\rho_m&-\delta_m&\rho_m&0\end{array}
\right)\lambda^{n-m}
+\Delta_n, \label{3.9} \eeq
with $\Delta_n$ being the modification term. We setting
$$\Delta_n=\left(\begin{array}{ccccc}
a&b&e&f&k\\
c&-a&g&e&l\\
0&0&a&b&0\\
0&0&c&-a&0\\
l&-k&-l&k&0\end{array}
\right),$$
and substitute Eq.\eqref{3.1} and Eq.\eqref{3.9} into the following zero curvature equation
\beq M_{t_n}-N_x^{(n)}+[M,N^{(n)}]=0, \label{3.10} \eeq
where $n\geq0$. Making use of Eq.\eqref{3.4} yields
\eqa
\left\{\begin{array}{l}
h_{t_n}=a_x,b=c=e=f=g=k=l=0,\\
p_{t_n}=b_{n,x}+2pa_{n}+2\alpha\rho_{n}-2h b_{n}+2pa=2b_{n+1}+2pa,\\
q_{t_n}=c_{n,x}-2qa_{n}-2\beta\delta_{n}+2h c_{n}-2qa=-2c_{n+1}-2qa,\\
\alpha_{t_n}=\rho_{n,x}+\alpha a_{n}+\beta b_{n}-p\delta_{n}-h \rho_{n}+\alpha a=\rho_{n+1}+\alpha a,\\
\beta_{t_n}=\delta_{n,x}-\beta a_{n}+\alpha c_{n}-q\rho_{n}+h \delta_{n}-\beta a=-\delta_{n+1}-\beta a,\\
r_{t_n}=f_{n,x}+2ra_{n}-2\alpha\rho_{n}-2(p+r)e_{n}-2h f_{n}+2ra=2f_{n+1}+2ra,\\
s_{t_n}=g_{n,x}-2sa_{n}+2\beta\delta_{n}-2(q+s)e_{n}+2h g_{n}-2sa=-2g_{n+1}-2sa,
\end{array}\right.\label{3.11}\eeqa
which guarantees the following identity:
\eqa &&
(ps+qr+rs-2\alpha\beta)_{t_n}=-2(rc_{n+1}-sb_{n+1}+(p+r)g_{n+1}-(q+s)f_{n+1}\nn\\&&\quad\quad\quad\quad\quad\quad\quad\quad\quad\quad\quad\;
-\alpha\delta_{n+1}-\beta\rho_{n+1})
=-2e_{n+1,x}.
\label{3.12} \eeqa
Choosing $a=-2\mu e_{n+1}$, we can obtain the following hierarchy:
\beq
{\small u_{t_n}=\left(\begin{array}{c}
p\\
q\\
\alpha\\
\beta\\
r\\
s\end{array} \right)_{t_n}=\left(\begin{array}{c}
2b_{n+1}-4\mu pe_{n+1}\\
-2c_{n+1}+4\mu qe_{n+1}\\
\rho_{n+1}-2\mu\alpha e_{n+1}\\
-\delta_{n+1}+2\mu\beta e_{n+1}\\
2f_{n+1}-4\mu re_{n+1}\\
-2g_{n+1}+4\mu se_{n+1}
\end{array} \right),n\geq0}. \label{3.13} \eeq

When $n=2$ in Eq.\eqref{3.13}, we obtain the first non-trivial flow as follows
 \beq  \left\{
\begin{array}{l}
p_{t_2}=\frac{1}{2}p_{xx}-p^2q-2p\alpha\beta+2\alpha\alpha_x+\mu(pp_xq-pp_xs-3prq_x-3prs_x+pr_xq+pr_xs\\\quad\quad\;
-2p_xqr-2p_xsr-p^2q_x-3p^2s_x+4p\alpha\beta_x-4p\alpha_x\beta+4p_x\alpha\beta)-2\mu^2p(2p^2sq+2pq^2r\\\quad\quad\;
+3q^2r^2+3s^2r^2+3p^2s^2+6ps^2r+6sqr^2+8psqr)+16\mu^2p\alpha\beta(ps+qr+sr+\frac{1}{2}pq),\\
q_{t_2}=-\frac{1}{2}q_{xx}+pq^2+2q\alpha\beta+2\beta\beta_x+\mu(pq_xq+pqs_x-3qsp_x-3qsr_x+qrq_x+qrs_x\\\quad\quad\;
-2psq_x-2srq_x-q^2p_x-3q^2r_x-4q\alpha\beta_x+4q\alpha_x\beta+4q_x\alpha\beta)+2\mu^2q(2p^2sq+2pq^2r\\\quad\quad\;
+3q^2r^2+3s^2r^2+3p^2s^2+6ps^2r+6sqr^2+8psqr)-16\mu^2q\alpha\beta(ps+qr+sr+\frac{1}{2}pq),\\
\alpha_{t_2}=\alpha_{xx}+\frac{1}{2}\alpha q_x+q\alpha_x-\frac{1}{2}pq\alpha
+\mu(\frac{1}{2}\alpha qp_x-\frac{1}{2}\alpha pq_x-2\alpha ps_x-2\alpha rs_x\\\quad\quad\;
-2\alpha rq_x-2ps\alpha_x-2qr\alpha_x-2sr\alpha_x+2\alpha\alpha_x\beta)-\mu^2\alpha(2p^2sq+2pq^2r\\\quad\quad\;
+3q^2r^2+3s^2r^2+3p^2s^2+6ps^2r+6sqr^2+8psqr),\\
\beta_{t_2}=-\beta_{xx}-\frac{1}{2}p_x\beta-p\beta_x+\frac{1}{2}pq\beta+\mu(\frac{1}{2}\beta pq_x-\frac{1}{2}\beta qp_x-2\beta sp_x-2\beta sr_x\\\quad\quad\;
-2\beta qr_x-2ps\beta_x-2qr\beta_x-2sr\beta_x+2\alpha\beta_x\beta)+\mu^2\beta(2p^2sq+2pq^2r\\\quad\quad\;
+3q^2r^2+3s^2r^2+3p^2s^2+6ps^2r+6sqr^2+8psqr),\\
r_{t_2}=r_{xx}+\frac{1}{2}p_{xx}-p^2q+2p\alpha\beta-2\alpha\alpha_x-2r^2s-2qr^2-2p^2s-4psr-4pqr+\mu(-3psp_x\\\quad\quad\;
-2prq_x-pqr_x-qrp_x-5prs_x-5psr_x-2srp_x-4qrr_x-4srr_x-4r^2q_x\\\quad\quad\;
-4r^2s_x-p^2s_x+4p_x\alpha\beta+4r\alpha\beta_x-4r\alpha_x\beta+4r_x\alpha\beta)+8\mu^2\alpha\beta(r^2q+r^2s-p^2s)\\\quad\quad\;
-2\mu^2(pq^2r^2-p^3s^2+2r^3q^2+2r^3s^2+3ps^2r^2+4r^3qs+4r^2psq),\\
s_{t_2}=-s_{xx}-\frac{1}{2}q_{xx}+pq^2-2q\alpha\beta-2\beta\beta_x+2rs^2+2rq^2+2ps^2+4pqs+4qrs\mu(-3qrq_x\\\quad\quad\;
-2srq_x-pqs_x-psq_x-5qrs_x-5qsr_x-2sqp_x-4pss_x-4rss_x-4s^2p_x\\\quad\quad\;
-4s^2r_x-q^2r_x+4q_x\alpha\beta-4s\alpha\beta_x+4s\alpha_x\beta+8s_x\alpha\beta)-8\mu^2\alpha\beta(r^2s+p^2s-r^2q)\\\quad\quad\;
+2\mu^2(qp^2s^2-q^3r^2+2s^3r^2+2s^3p^2+3qs^2r^2+4s^3pr+4s^2prq),
\end{array}
\right. \label{3.14} \eeq
whose Lax pair consists of $M$ and $N^{(2)}$. $M$ is defined by \eqref{3.1} and $N^{(2)}$
\beq
N^{(2)}=\left(\begin{array}{ccccc}
N_{11}^{(2)} & N_{12}^{(2)} & N_{13}^{(2)} & N_{14}^{(2)} & N_{15}^{(2)}\\
N_{21}^{(2)} & -N_{11}^{(2)} & N_{23}^{(2)} & -N_{13}^{(2)} & N_{25}^{(2)}\\
0&0&N_{33}^{(2)} & N_{34}^{(2)} &0\\
0&0&N_{43}^{(2)} & -N_{33}^{(2)} &0\\
N_{25}^{(2)}&-N_{15}^{(2)}&-N_{25}^{(2)}&N_{15}^{(2)}&0
\end{array} \right).\label{3.15}\eeq
with
\eqa &&
N_{11}^{(2)}=\lambda^2-\frac{1}{2}(pq+2\alpha\beta),
N_{12}^{(2)}=p\lambda+\frac{1}{2}p_x-hp,
N_{13}^{(2)}=\lambda^2-(ps+qr+rs+\frac{1}{2}pq-\alpha\beta),\nn\\&&
N_{14}^{(2)}=(p+2r)\lambda+\frac{1}{2}p_x+\frac{1}{2}r_x-hp-hr,
N_{15}^{(2)}=\alpha\lambda+\frac{1}{2}\alpha_x-h\alpha,
N_{21}^{(2)}=q\lambda-\frac{1}{2}q_x-hq,\nn\\&&
N_{23}^{(2)}=(q+2s)\lambda-\frac{1}{2}q_x-\frac{1}{2}s_x-hq-hs,
N_{25}^{(2)}=\beta\lambda-\frac{1}{2}\beta_x-h\beta,\nn\\&&
N_{33}^{(2)}=2\lambda^2-(ps+qr+rs+pq),
N_{34}^{(2)}=2(p+r)\lambda+p_x+\frac{1}{2}r_x-2hp-hr,\nn\\&&
N_{43}^{(2)}=2(q+s)\lambda-q_x-\frac{1}{2}s_x-2hq-hs. \nn\eeqa

\section{Super bi-Hamiltonian structures}

\quad\;\;In what follows we shall find super bi-Hamiltonian structures of the
nonlinear super integrable couplings of a generalized super AKNS hirearchy \eqref{3.13}.
To this end, we shall apply the super variational identities, which was discussed in \cite{MA2013}
\beq \frac{\delta}{\delta u}\int Str(N\frac{\p M}{\p \lambda})dx=(\lambda^{-\gamma}\frac{\p}{\p\lambda}\lambda^{\gamma})Str(\frac{\p M}{\p u}N),\label{4.1}\eeq
where $Str$ denotes the super trace. It is not difficult to find that
\eqa &&
Str(N\frac{\p M}{\p \lambda})=4A+2E,
Str(\frac{\p M}{\p p}N)=2C+G+2\mu s(2A+E),\nn\\&&
Str(\frac{\p M}{\p q}N)=2B+F+2\mu r(2A+E),
Str(\frac{\p M}{\p \alpha}N)=2\delta+4\mu\beta(2A+E),\nn\\&&
Str(\frac{\p M}{\p \beta}N)=-2\rho-4\mu\alpha(2A+E),
Str(\frac{\p M}{\p r}N)=C+G+2\mu(q+s)(2A+E),\nn\\&&
Str(\frac{\p M}{\p s}N)=B+F+2\mu(p+r)(2A+E).
\label{4.2}\eeqa
Substituting Eq.\eqref{4.2} into Eq.\eqref{4.1}, and comparing the coefficient of $\lambda^{-n-2}$ of both sides of Eq.\eqref{4.1} yields
\beq \left(\begin{array}{c}
\frac{\delta}{\delta p}\\
\frac{\delta}{\delta q}\\
\frac{\delta}{\delta \alpha}\\
\frac{\delta}{\delta \beta}\\
\frac{\delta}{\delta r}\\
\frac{\delta}{\delta s}
\end{array}
\right)\int(4a_{n+2}+2e_{n+2})dx=(\gamma-n-1)\left(\begin{array}{c}
2c_{n+1}+g_{n+1}+2\mu s(2a_{n+1}+e_{n+1})\\
2b_{n+1}+f_{n+1}+2\mu r(2a_{n+1}+e_{n+1})\\
2\delta_{n+1}+4\mu\beta(2a_{n+1}+e_{n+1})\\
-2\rho_{n+1}-4\mu\alpha(2a_{n+1}+e_{n+1})\\
c_{n+1}+g_{n+1}+2\mu(q+s)(2a_{n+1}+e_{n+1})\\
b_{n+1}+f_{n+1}+2\mu(p+r)(2a_{n+1}+e_{n+1})
\end{array} \right). \label{4.3}\eeq
The identity with $n=0$ tells $\gamma=0$. Thus, we have
\beq
\left(\begin{array}{c}
2c_{n+1}+g_{n+1}+2\mu s(2a_{n+1}+e_{n+1})\\
2b_{n+1}+f_{n+1}+2\mu r(2a_{n+1}+e_{n+1})\\
2\delta_{n+1}+4\mu\beta(2a_{n+1}+e_{n+1})\\
-2\rho_{n+1}-4\mu\alpha(2a_{n+1}+e_{n+1})\\
c_{n+1}+g_{n+1}+2\mu(q+s)(2a_{n+1}+e_{n+1})\\
b_{n+1}+f_{n+1}+2\mu(p+r)(2a_{n+1}+e_{n+1})
\end{array} \right)
=\frac{\delta \tilde{H}_{n+1}}{\delta u}, \label{4.4}\eeq
where $\tilde{H}_{n+1}=-2\int\frac{2a_{n+2}+e_{n+2}}{n+1}dx$. Moreover, a direct calculation yields to the following recursive relationship
\beq
\left(\begin{array}{c}
c_{n+1}\\
b_{n+1}\\
\delta_{n+1}\\
\rho_{n+1}\\
g_{n+1}\\
f_{n+1}
\end{array} \right)
=R\left(\begin{array}{c}
2c_{n+1}+g_{n+1}+2\mu s(2a_{n+1}+e_{n+1})\\
2b_{n+1}+f_{n+1}+2\mu r(2a_{n+1}+e_{n+1})\\
2\delta_{n+1}+4\mu\beta(2a_{n+1}+e_{n+1})\\
-2\rho_{n+1}-4\mu\alpha(2a_{n+1}+e_{n+1})\\
c_{n+1}+g_{n+1}+2\mu(q+s)(2a_{n+1}+e_{n+1})\\
b_{n+1}+f_{n+1}+2\mu(p+r)(2a_{n+1}+e_{n+1})
\end{array} \right)
\label{4.5}\eeq
where $R$ is given by
\beq
R=\left(\begin{array}{ccc}
R_{11} & R_{12} & R_{13}\\
R_{21} & R_{22} & R_{23}\\
R_{31} & R_{32} & R_{33}
\end{array} \right)
\nn\eeq
with
$$
R_{11}=\left(\begin{array}{cc}
1+2\mu q\p^{-1}p & -2\mu q\p^{-1}q    \\
 2\mu p\p^{-1}p & 1-2\mu p\p^{-1}q
\end{array} \right),
R_{12}=\left(\begin{array}{cc}
 \mu q\p^{-1}\alpha  &  -\mu q\p^{-1}\beta   \\
\mu p\p^{-1}\alpha  &  -\mu p\p^{-1}\beta
\end{array} \right),$$
$$R_{13}=\left(\begin{array}{cc}
-1+2\mu q\p^{-1}r  & -2\mu q\p^{-1}s    \\
 2\mu p\p^{-1}r &   -1-2\mu p\p^{-1}s
\end{array} \right),
R_{21}=\left(\begin{array}{cc}
-2\mu\beta\p^{-1}p & 2\mu\beta\p^{-1}q    \\
-2\mu\alpha\p^{-1}p & 2\mu\alpha\p^{-1}q
\end{array} \right),$$
$$R_{22}=\left(\begin{array}{cc}
\frac{1}{2}-\mu\beta\p^{-1}\alpha  & \mu\beta\p^{-1}\beta    \\
-\mu\alpha\p^{-1}\alpha & -\frac{1}{2}+\mu\alpha\p^{-1}\beta
\end{array} \right),
R_{23}=\left(\begin{array}{cc}
-2\mu\beta\p^{-1}r  & 2\mu\beta\p^{-1}s    \\
-2\mu\alpha\p^{-1}r & 2\mu\alpha\p^{-1}s
\end{array} \right).$$
$$R_{31}=\left(\begin{array}{cc}
-1-2\mu(2q+s)\p^{-1}p & 2\mu(2q+s)\p^{-1}q  \\
-2\mu(2p+r)\p^{-1}p & -1+2\mu(2p+r)\p^{-1}q
\end{array} \right),$$
$$R_{32}=\left(\begin{array}{cc}
-\mu(2q+s)\p^{-1}\alpha  &  \mu(2q+s)\p^{-1}\beta\\
-\mu(2p+r)\p^{-1}\alpha  &  \mu(2p+r)\p^{-1}\beta
\end{array} \right),$$
$$R_{33}=\left(\begin{array}{cc}
2-2\mu(2q+s)\p^{-1}r &  2\mu(2q+s)\p^{-1}s  \\
 -2\mu(2p+r)\p^{-1}r  & 2+2\mu(2p+r)\p^{-1}s
\end{array} \right).$$

Thus, the hierarchy \eqref{3.13} possesses the following
super-Hamiltonian structure
\beq
{\small u_{t_n}=Q\left(\begin{array}{c}
c_{n+1}\\
b_{n+1}\\
\delta_{n+1}\\
\rho_{n+1}\\
g_{n+1}\\
f_{n+1}
\end{array} \right)=QR\left(\begin{array}{c}
2c_{n+1}+g_{n+1}+2\mu s(2a_{n+1}+e_{n+1})\\
2b_{n+1}+f_{n+1}+2\mu r(2a_{n+1}+e_{n+1})\\
2\delta_{n+1}+4\mu\beta(2a_{n+1}+e_{n+1})\\
-2\rho_{n+1}-4\mu\alpha(2a_{n+1}+e_{n+1})\\
c_{n+1}+g_{n+1}+2\mu(q+s)(2a_{n+1}+e_{n+1})\\
b_{n+1}+f_{n+1}+2\mu(p+r)(2a_{n+1}+e_{n+1})
\end{array} \right)}=J\frac{\delta \tilde{H}_n}{\delta u}, \label{4.6} \eeq
where
\beq
Q=\left(\begin{array}{ccc}
Q_{11} & Q_{12} & Q_{13}\\
Q_{21} & Q_{22} & Q_{23}\\
Q_{31} & Q_{32} & Q_{33}
\end{array} \right)
\nn\eeq
with
$$
Q_{11}=\left(\begin{array}{cc}
-4\mu p\p^{-1}r & 2+4\mu p\p^{-1}s    \\
-2+4\mu q\p^{-1}r & -4\mu q\p^{-1}s
\end{array} \right),
Q_{12}=\left(\begin{array}{cc}
 4\mu p\p^{-1}\alpha  &  4\mu p\p^{-1}\beta   \\
-4\mu q\p^{-1}\alpha  &  -4\mu q\p^{-1}\beta
\end{array} \right),$$
$$Q_{13}=\left(\begin{array}{cc}
 -4\mu p\p^{-1}(p+r) & 4\mu p\p^{-1}(q+s) \\
 4\mu q\p^{-1}(p+r)& -4\mu q\p^{-1}(q+s)
\end{array} \right),
Q_{21}=\left(\begin{array}{cc}
-2\mu\alpha\p^{-1}r & 2\mu\alpha\p^{-1}s    \\
2\mu\beta\p^{-1}r & -2\mu\beta\p^{-1}s
\end{array} \right),$$
$$Q_{22}=\left(\begin{array}{cc}
2\mu \alpha\p^{-1}\alpha  &  1+2\mu \alpha\p^{-1}\beta   \\
-1-2\mu \beta\p^{-1}\alpha  &  -2\mu \beta\p^{-1}\beta
\end{array} \right)$$
$$Q_{23}=\left(\begin{array}{cc}
 -2\mu \alpha\p^{-1}(p+r) & 2\mu \alpha\p^{-1}(q+s)  \\
 2\mu \beta\p^{-1}(p+r)& -2\mu \beta\p^{-1}(q+s)
\end{array} \right),
Q_{31}=\left(\begin{array}{cc}
-4\mu r\p^{-1}r & 4\mu r\p^{-1}s    \\
4\mu s\p^{-1}r & -4\mu s\p^{-1}s
\end{array} \right),$$
$$Q_{32}=\left(\begin{array}{cc}
4\mu r\p^{-1}\alpha  &  4\mu r\p^{-1}\beta   \\
-4\mu s\p^{-1}\alpha  &  -4\mu s\p^{-1}\beta
\end{array} \right),
Q_{33}=\left(\begin{array}{cc}
-4\mu r\p^{-1}(p+r) & 2+4\mu r\p^{-1}(q+s)  \\
-2+4\mu s\p^{-1}(p+r) & -4\mu s\p^{-1}(q+s)
\end{array} \right)
.$$
and
\beq
J=QR=\left(\begin{array}{ccc}
J_{1} & J_{2} & -J_{1}\\
0 & J_{3} & J_{4}\\
-J_{1} & -J_{2} & J_{5}
\end{array} \right)
\nn\eeq
with
$$
J_{1}=\left(\begin{array}{cc}
8\mu p\p^{-1}p & 2-4\mu p\p^{-1}q    \\
-2-8\mu q\p^{-1}p & 8\mu q\p^{-1}q
\end{array} \right),
J_{2}=\left(\begin{array}{ccc}
 4\mu p\p^{-1}\alpha  &  -4\mu p\p^{-1}\beta   \\
-4\mu q\p^{-1}\alpha  &  4\mu q\p^{-1}\beta
\end{array} \right),$$
$$J_{3}=\left(\begin{array}{cc}
0 & -\frac{1}{2}    \\
-\frac{1}{2}  & 0
\end{array} \right),
J_{4}=\left(\begin{array}{cc}
-4\mu\alpha\p^{-1}(p+r) & 4\mu\alpha\p^{-1}(q+s)    \\
4\mu\beta\p^{-1}(p+r) & -4\mu\beta\p^{-1}(q+s)
\end{array} \right),$$
$$J_{5}=\left(\begin{array}{cc}
-8\mu r\p^{-1}(p+r)-8\mu p\p^{-1}r & 4+8\mu r\p^{-1}(q+s)+8\mu p\p^{-1}s  \\
-4+8\mu s\p^{-1}(p+r)+8\mu q\p^{-1}r & -8\mu s\p^{-1}(q+s)-8\mu q\p^{-1}s
\end{array} \right).$$\\
It could be proved that $J$ is a super Hamiltonian operator.

Specially, by making use of the recursive relationship \eqref{3.7},
the hierarchy \eqref{3.13} possesses the following
super-bi-Hamiltonian structure
\beq
{\small u_{t_n}=QL\left(\begin{array}{c}
c_{n}\\
b_{n}\\
\delta_{n}\\
\rho_{n}\\
g_{n}\\
f_{n}
\end{array} \right)=QLR\left(\begin{array}{c}
2c_{n}+g_{n}+2\mu s(2a_{n}+e_{n})\\
2b_{n}+f_{n}+2\mu r(2a_{n}+e_{n})\\
2\delta_{n}+4\mu\beta(2a_{n}+e_{n})\\
-2\rho_{n}-4\mu\alpha(2a_{n}+e_{n})\\
c_{n}+g_{n}+2\mu(q+s)(2a_{n}+e_{n})\\
b_{n}+f_{n}+2\mu(p+r)(2a_{n}+e_{n})
\end{array} \right)}=P\frac{\delta \tilde{H}_{n-1}}{\delta u}, n\geq2. \label{4.7} \eeq
where the second compatible super-Hamiltonian operator $P=QLR=(P_{ij})_{6\times 6}, i,j=1,2,\ldots,6,$ is given by
\eqa &&
P_{11}=2p\p^{-1}p-4\mu p\p^{-1}p(\frac{1}{2}\p+h)+2\mu(\p-2h)p\p^{-1}p-4\mu^2p\Delta\p^{-1}p,\nn\\&&
P_{12}=-2p\p^{-1}q-4\mu p\p^{-1}q(\frac{1}{2}\p-h)-2\mu(\p-2h)p\p^{-1}q+4\mu^2p\Delta\p^{-1}q,\nn\\&&
P_{13}=p\p^{-1}\alpha-2\mu p\p^{-1}\alpha(\p+h)+\mu(\p-2h)p\p^{-1}\alpha-2\mu^2p\Delta\p^{-1}\alpha,\nn\\&&
P_{14}=-\alpha-p\p^{-1}\beta-2\mu p\p^{-1}\beta(\p-h)-\mu(\p-2h)p\p^{-1}\beta+2\mu^2p\Delta\p^{-1}\beta,\nn\\&&
P_{15}=-2p\p^{-1}p+4\mu p\p^{-1}(2p+r)(\frac{1}{2}\p+h)+2\mu(\p-2h)p\p^{-1}r-4\mu^2p\Delta\p^{-1}r,\nn\\&&
P_{16}=2p\p^{-1}q+4\mu p\p^{-1}(2q+s)(\frac{1}{2}\p-h)-2\mu(\p-2h)p\p^{-1}s+4\mu^2p\Delta\p^{-1}s,\nn\\&&
P_{21}=-2q\p^{-1}p+4\mu q\p^{-1}p(\frac{1}{2}\p+h)+2\mu(\p+2h)q\p^{-1}p+4\mu^2q\Delta\p^{-1}p,\nn\\&&
P_{22}=2q\p^{-1}q+4\mu q\p^{-1}q(\frac{1}{2}\p-h)-2\mu(\p+2h)q\p^{-1}q-4\mu^2q\Delta\p^{-1}q,\nn\\&&
P_{23}=-\beta-q\p^{-1}\alpha+2\mu q\p^{-1}\alpha(\p+h)+\mu(\p+2h)q\p^{-1}\alpha+2\mu^2q\Delta\p^{-1}\alpha,\nn\\&&
P_{24}=q\p^{-1}\beta+2\mu q\p^{-1}\beta(\p-h)-\mu(\p+2h)q\p^{-1}\beta-2\mu^2q\Delta\p^{-1}\beta,\nn\\&&
P_{25}=2q\p^{-1}p-4\mu q\p^{-1}(2p+r)(\frac{1}{2}\p+h)+2\mu(\p+2h)q\p^{-1}r+4\mu^2q\Delta\p^{-1}r,\nn\\&&
P_{26}=-2q\p^{-1}q-4\mu q\p^{-1}(2q+s)(\frac{1}{2}\p-h)-2\mu(\p+2h)q\p^{-1}s-4\mu^2q\Delta\p^{-1}s,\nn\\&&
P_{31}=\alpha\p^{-1}p-2\mu \alpha\p^{-1}p(\frac{1}{2}\p+h)+4\mu\beta p\p^{-1}p-2\mu(\p-h)\alpha\p^{-1}p-2\mu^2\alpha\Delta\p^{-1}p,\nn\\&&
P_{32}=\beta-\alpha\p^{-1}q-2\mu \alpha\p^{-1}q(\frac{1}{2}\p-h)-4\mu\beta p\p^{-1}q+2\mu(\p-h)\alpha\p^{-1}q+2\mu^2\alpha\Delta\p^{-1}q,\nn\\&&
P_{33}=-\frac{1}{2}p+\frac{1}{2}\alpha\p^{-1}\alpha-\mu \alpha\p^{-1}\alpha(\p+h)
+2\mu\beta p\p^{-1}\alpha-\mu(\p-h)\alpha\p^{-1}\alpha-\mu^2\alpha\Delta\p^{-1}\alpha,\nn\\&&
P_{34}=\frac{1}{2}h-\frac{1}{2}\p-\frac{1}{2}\alpha\p^{-1}\beta-\mu \alpha\p^{-1}\beta(\p-h)
-2\mu\beta p\p^{-1}\beta+\mu(\p-h)\alpha\p^{-1}\beta+\mu^2\alpha\Delta\p^{-1}\beta,\nn\\&&
P_{35}=-\alpha\p^{-1}p+2\mu \alpha\p^{-1}(2p+r)(\frac{1}{2}\p+h)+4\mu\beta p\p^{-1}r-2\mu(\p-h)\alpha\p^{-1}r-2\mu^2\alpha\Delta\p^{-1}r,\nn\\&&
P_{36}=-\beta+\alpha\p^{-1}q+2\mu \alpha\p^{-1}(2q+s)(\frac{1}{2}\p-h)-4\mu\beta p\p^{-1}s+2\mu(\p-h)\alpha\p^{-1}s+2\mu^2\alpha\Delta\p^{-1}s,\nn\\&&
P_{41}=\alpha-\beta\p^{-1}p+2\mu \beta\p^{-1}p(\frac{1}{2}\p+h)+4\mu\alpha q\p^{-1}p-2\mu(\p+h)\beta\p^{-1}p+2\mu^2\beta\Delta\p^{-1}p,\nn\\&&
P_{42}=\beta\p^{-1}q+2\mu \beta\p^{-1}q(\frac{1}{2}\p-h)-4\mu\alpha q\p^{-1}q+2\mu(\p+h)\beta\p^{-1}q-2\mu^2\beta\Delta\p^{-1}q,\nn\\&&
P_{43}=\frac{1}{2}h+\frac{1}{2}\p-\frac{1}{2}\beta\p^{-1}\alpha+\mu\beta\p^{-1}\alpha(\p+h)
+2\mu\alpha q\p^{-1}\alpha-\mu(\p+h)\beta\p^{-1}\alpha+\mu^2\beta\Delta\p^{-1}\alpha,\nn\\&&
P_{44}=\frac{1}{2}q+\frac{1}{2}\beta\p^{-1}\beta+\mu\beta\p^{-1}\beta(\p-h)
-2\mu\alpha q\p^{-1}\beta+\mu(\p+h)\beta\p^{-1}\beta-\mu^2\beta\Delta\p^{-1}\beta,\nn\\&&
P_{45}=-\alpha+\beta\p^{-1}p-2\mu \beta\p^{-1}(2p+r)(\frac{1}{2}\p+h)+4\mu\alpha q\p^{-1}r-2\mu(\p+h)\beta\p^{-1}r+2\mu^2\beta\Delta\p^{-1}r,\nn\\&&
P_{46}=-\beta\p^{-1}q-2\mu \beta\p^{-1}(2q+s)(\frac{1}{2}\p-h)-4\mu\alpha q\p^{-1}s+2\mu(\p+h)\beta\p^{-1}s-2\mu^2\beta\Delta\p^{-1}s,\nn\\&&
P_{51}=-2p\p^{-1}p-4\mu r\p^{-1}p(\frac{1}{2}\p+h)-2\mu(\p-2h)[(2p+r)\p^{-1}p]-4\mu^2r\Delta\p^{-1}p,\nn\\&&
P_{52}=-\p+2h+2p\p^{-1}q-4\mu r\p^{-1}q(\frac{1}{2}\p-h)+2\mu(\p-2h)[(2p+r)\p^{-1}q]+4\mu^2r\Delta\p^{-1}q,\nn\\&&
P_{53}=-p\p^{-1}\alpha-2\mu r\p^{-1}\alpha(\p+h)-\mu(\p-2h)[(2p+r)\p^{-1}\alpha]-2\mu^2r\Delta\p^{-1}\alpha,\nn\\&&
P_{54}=p\p^{-1}\beta-2\mu r\p^{-1}\beta(\p-h)+\mu(\p-2h)[(2p+r)\p^{-1}\beta]+2\mu^2r\Delta\p^{-1}\beta,\nn\\&&
P_{55}=2(2p+r)\p^{-1}p+2(p+r)\p^{-1}r+4\mu r\p^{-1}(2p+r)(\frac{1}{2}\p+h)\nn\\&&\quad\quad\quad
-2\mu(\p-2h)[(2p+r)\p^{-1}r]-4\mu^2r\Delta\p^{-1}r,\nn\\&&
P_{56}=2\p-4h+2(2q+s)\p^{-1}q+2(q+s)\p^{-1}s+4\mu r\p^{-1}(2q+s)(\frac{1}{2}\p-h)\nn\\&&\quad\quad\quad
+2\mu(\p-2h)[(2p+r)\p^{-1}s]+4\mu^2r\Delta\p^{-1}s,\nn\\&&
P_{61}=-\p-2h+2q\p^{-1}p+4\mu s\p^{-1}p(\frac{1}{2}\p+h)-2\mu(\p+2h)[(2q+s)\p^{-1}p]+4\mu^2s\Delta\p^{-1}p,\nn\\&&
P_{62}=-2q\p^{-1}q-4\mu s\p^{-1}q(\frac{1}{2}\p-h)+2\mu(\p+2h)[(2q+s)\p^{-1}q]-4\mu^2s\Delta\p^{-1}q,\nn\\&&
P_{63}=-p\p^{-1}\alpha-2\mu s\p^{-1}\alpha(\p+h)-\mu(\p+2h)[(2q+s)\p^{-1}\alpha]+2\mu^2s\Delta\p^{-1}\alpha,\nn\\&&
P_{64}=p\p^{-1}\beta-2\mu s\p^{-1}\beta(\p-h)+\mu(\p+2h)[(2q+s)\p^{-1}\beta]-2\mu^2s\Delta\p^{-1}\beta,\nn\\&&
P_{65}=2\p+4h+2(2p+r)\p^{-1}p+2(p+r)\p^{-1}r+4\mu s\p^{-1}(2q+s)(\frac{1}{2}\p+h)\nn\\&&\quad\quad\quad
-2\mu(\p+2h)[(2q+s)\p^{-1}r]+4\mu^2s\Delta\p^{-1}r,\nn\\&&
P_{66}=2\p-4h+2(2q+s)\p^{-1}q+2(q+s)\p^{-1}s+4\mu s\p^{-1}(2q+s)(\frac{1}{2}\p-h)\nn\\&&\quad\quad\quad
+2\mu(\p+2h)[(2q+s)\p^{-1}s]-4\mu^2s\Delta\p^{-1}s
.\nn\eeqa
with
$$\Delta=\p^{-1}(2q+s)\p p+\p^{-1}(2p+r)\p q+\p^{-1}(q+s)\p r+\p^{-1}(p+r)\p s+2\p^{-1}\beta\p\alpha-2\p^{-1}\alpha\p\beta.$$

\section{Conclusion and discussions}

\quad\;\;In this paper, we presented an approach for constructing nonlinear
super integrable couplings of super soliton equations through enlarging matrix
Lie super algebras. We took the Lie algebra $sl(2,1)$ as an example to illustrate
the introduced idea to extend Lie super algebras. Based on the enlarged Lie super
algebra $sl(4,1)$, we worked out nonlinear integrable couplings for a generalized
super AKNS soliton hierarchy. The presented method in this paper can be applied to
other generalized super integrable hierarchies, which will be our future problems
to construct super integrable couplings.\\
\\
\textbf{Acknowledgements}\\

The work was supported by NSFC under the grants 11601055, 11371326, 11301331,
and 11371086, NSF under the grant DMS-1664561, the 111 project of China (B16002), and
the Distinguished Professorships by Shanghai University of Electric Power and Shanghai
Second Polytechnic University.


\begin{thebibliography}{99}
\itemsep=0cm
\small

\bibitem{wess1974}
Wess J, Zumino B. Supergauge transformations in four dimensions. Nucl. Phys.B 1974;70:39-50.

\bibitem{Gurses1985}
Gurses M, Oguz O. A super AKNS scheme. Phys Lett A 1985;108:437-440.
\bibitem{lzh1986}
Li YS, Zhang LN. Super AKNS scheme and its infinite conserved currents. Nuovo Cimento A 1986;93:175-183.
\bibitem{Ma2008}
Ma WX, He JS, Qin ZY. A supertrace identity and its applications to super integrable systems. J. Math. Phys 2008;49:033511.
\bibitem{He2008}
He JS, Yu J, Cheng Y, Zhou RG. Binary nonlinearization of the super AKNS system. Mod Phys Lett B 2008;22:275-288.
\bibitem{Yujing2009}
Yu J, Han JW, He JS. Binary nonlinearization of the super AKNS system under an implicit symmetry constraint. J Phys A 2009;42:465201.
\bibitem{yujing2014}	
Yu J, Han JW. Two-Component Super AKNS Equations and Their Finite-Dimensional Integrable Super Hamiltonian System. Abstract and Applied Analysis 2014;507540.
\bibitem{You2014}
You FC, Zhang J, Zhao Y. Super-Hamiltonian Structures and Conservation Laws of a New Six-Component Super-Ablowitz-Kaup-Newell-Segur Hierarchy. Abstract and Applied Analysis 2014;214709.
\bibitem{yujing2017}
Yu J, Ma WX, Han JW, Chen ST. An integrable generalization of the super AKNS hierarchy and its bi-Hamiltonian formulation. Commun Nonlinear Sci Numer Simulat. 2017;43:151-157.
\bibitem{Han2017}
Han JW, Yu J. A generalized super AKNS hierarchy associated with Lie superalgebra $sl(2|1)$ and its super bi-Hamiltonian structure. Commun Nonlinear Sci Numer Simulat 2017;44:258-265.


\bibitem{Yu2010}
Yu J, He JS, Ma WX, Cheng Y. The Bargmann Symmetry Constraint and Binary Nonlinearization of the Super Dirac Systems. Chin. Ann. Math 2010;31:361-372.
\bibitem{You2012}
You FC. Nonlinear Super Integrable Couplings of Super Dirac Hierarchy and Its Super Hamiltonian Structures. Commun. Theor. Phys 2012;57:961-966.
\bibitem{Zhangjiao2014}
Zhang J, You FC, Zhao Y. A New Super Extension of Dirac Hierarchy. Abstract and Applied Analysis Volume 2014;472101.
\bibitem{Ye2017}
Ye YJ, Li ZH, Shen SF, Li CX. A generalized super integrable hierarchy of Dirac type. arXiv:1604.03728


\bibitem{Geng2010}
Geng XG, Wu LH. A super extension of Kaup-Newell hierarchy. Commun Theor Phys 2010;54:594-598.
\bibitem{Tao2011}
Tao SX, Xia TC, Shi H. Super-KN hinarchy and its super-Hamiltonian structure. Comm. Theor. Phys 2011;55:391-395.
\bibitem{wang2013}
Wang H, Xia TC. Conservation laws and self-consistent sources for a super KN hierarchy. Appl. Math. Comput 2013;219:5458-5464.
\bibitem{HU2017}
Hu BB, Xia TC, Zhang L. An integrable generalization of the super Kaup-Newell soliton hierarchy and its bi-Hamiltonian structure. arXiv:1706.03929



\bibitem{shaw1998}
Shaw JC, Tu MH. Binary Darboux-B\"{a}cklund transformations for the Manin-Radul super KdV hierarchy. J Math Phys 1998;39:4773-4784.
\bibitem{Gomes2006}
Gomes JF, Ymai LH, Zimerman AH. Soliton solutions for the super mKdV and sinh-Gordon hierarchy. Phy Lett A 2006;359:630-637.
\bibitem{Belitsky2008}
Belitsky AV. Fusion hierarchies for $n=4$ super-Yang-Mills theorey. Nucl Phys B 2008;803:171-193.
\bibitem{Aratyn2008}
Aratyn H, Gomes JF, Ymai LH, Zimerman AH. A class of soliton solutions for the $n=2$ super mKdV/sinh-Gordon hierarchy. J Phys A 2008;41:312001.
\bibitem{tx2010}
Tao SX, Xia TC. Lie algebra and Lie super algebra for integrable couplingof C-KdV hierarchy. Chin Phys Lett. 2010;27:040202.
\bibitem{tx2011}
Tao SX, Xia TC. Two super-integrable hierarchies and their super-Hamiltonian structures, Commun Nonlinear Sci Numer Simulat 2011;16:127-132.
\bibitem{Dong2015}
Dong HH, Zhao K, Yang HW, Li YQ. Generalised (2+1)-Dimensional Super Mkdv Hierarchy For Integrable Systems In Soliton Theory. East Asian Journal on Applied Mathematics 2015;5:256-72.
\bibitem{hu2015}
Hu BB, Zhang L, Fang F. Super-Li Spectrum Problems and Its Self-consistent Source. Journal of Jilin University(Science Edition) 2015;53:229-234.
\bibitem{Hu2017}
Hu BB, Xia TC. The Binary Nonlinearization of the Super
Integrable System and Its Self-Consistent Sources. International Journal of Nonlinear Sciences and Numerical Simulation 2017;18:285-292.
\bibitem{Yang2010}
Yang HX, Du J, Xu XX, Cui JP. Hamiltonian and Super-Hamiltonian Systems of a Hierarchy of Soliton Equations. Applied Mathematics and Computation 2010;217: 1497-1508.

\bibitem{hu1997}
Hu XB. An approach to generate super extensions of integrable systems. J. Phys. A 1997;30:619-632.
\bibitem{ma2008}
Ma WX, He JS, Qin ZY. A super trace identity and its applications to super integrable systems. J. Math. Phys 2008;49:033511.
\bibitem{ma2010}
Ma WX. Variational identities and Hamiltonian structures/W. X. Ma, X. B. Hu and Q. P. Liu, Nonlinear and Modern Mathematical Phasics. Melville, NY: American Institute of Physics 2010;1-27.


\bibitem{Tao2013}
Tao SX, Xia TC. Nonlinear Super Integrable Couplings of Super Broer-Kaup-Kupershmidt Hierarchy and Its Super Hamiltonian Structures. Advances in Mathematical Physics 2013;520765.
\bibitem{Wei2013}
Wei HY, Xia TC. Nonlinear integrable couplings of super Kaup-Newell hierarchy and its super Hamiltonian structures. Acta Phys. Sin. 2013;62:120202.
\bibitem{You2011}
You FC. Nonlinear super integrable Hamiltonian couplings. Journal of Mathematical Physics 2011;52:123510.
\bibitem{You2012}
You FC. Nonlinear Super Integrable Couplings of Super Dirac Hierarchy and Its Super Hamiltonian Structures. Commun. Theor. Phys. 2012;57:961-966.
\bibitem{You2015}
You FC, Zhang J. Nonlinear Superintegrable Couplings for Supercoupled KDV Hierarchy with Self-Consistent Sources. Reports on Mathematical Physics  2015;76:131-140.
\bibitem{Xing2014}
Xing XZ, Wu JZ, Geng XG. Nonlinear Super Integrable Couplings of Super Classical-Boussinesq Hierarchy. Journal of Applied Mathematics 2014;438741.
\bibitem{Chen2014}
Chen XH, Zhang HQ, You FC, Zhang DQ. A Super Integrable Hierarchy and Its Nonlinear Super Integrable Hamiltonian Couplings. Reports on Mathematical Physics 2014;74:205-220.

\bibitem{MA2010}
Ma WX, Zhu ZN. Constructing nonlinear discrete integrable Hamiltonian couplings, Computers and Mathematics with Applications 2010;60:2601-2608.
\bibitem{MA2011}
Ma WX. Nonlinear continuous integrable Hamiltonian couplings, Applied Mathematics and Computation 2011;217:7238-7244.
\bibitem{MA2013}
Ma WX, Meng JH, Zhang HQ. Tri-integrable couplings by matrix loop algebras, International Journal of Nonlinear Sciences and Numerical Simulation 2013;14:377-388.
\bibitem{Zhang2011}
Zhang YF. Lie algebras for constructing nonlinear integrable couplings. Communications in Theoretical Physics 2011;56:805-812.
\bibitem{Zhang2012}
Zhang YF, Mei JQ. Lie algebras and integrable systems. Communications in Theoretical Physics, 2012;57:1012-1022.
\bibitem{LI2014}
Li YQ, Dong HH, Yin BS. A Hierarchy of Discrete Integrable Coupling System With Self-Consistent Sources. Journal of Applied Mathematics, 2014;416472.
\bibitem{XU2010}
Xu XX. An Integrable Coupling Hierarchy of The Mkdv Integrable Systems, Its Hamiltonian Structure and Corresponding Nonisospectral Integrable Hierarchy. Applied Mathematics and Computation 2010;216:344-353.

\bibitem{Shen2017}
Shen SF, Li CX, Jin YY, Ma WX. Completion of the Ablowitz-Kaup-Newell-Segur integrable coupling.  arXiv:1706.04308
\bibitem{Shen2014}
Shen SF, Li CX, Jin YY, Yu SM. Multi-component integrable couplings for the Ablowitz-Kaup-Newell-Segur and Volterra hierarchies. Mathematical Methods in the Applied Sciences 2015;38:4345-4356.

\bibitem{Mawenxiu2013}
Ma WX.Integrable couplings and matrix loop algebras. Nonlinear and Modern Mathematical Physics, 2013;1562;105-122.

\end{thebibliography}
\end{document}